\documentclass{article}
\usepackage{spconf,amsmath,amssymb,graphicx,booktabs}
\usepackage[table]{xcolor}
\usepackage{tikz,microtype,cite,url,hyperref}
\usepackage{pgfplots,float}
\usetikzlibrary{arrows.meta, backgrounds, fit}
\pgfplotsset{compat=1.18}
\usetikzlibrary{arrows.meta,positioning,calc}
\hypersetup{hidelinks,pdftitle={Replacing Target Construction with Representation Regularization in Speech SSL}}
\definecolor{MethodBlue}{RGB}{225,237,248}
\definecolor{MethodBorder}{RGB}{50,95,135}
\definecolor{OursGreen}{RGB}{224,242,229}
\newcommand{\method}{\textup{GLaS-JEPA}}
\newcommand{\sg}{\operatorname{sg}}

\definecolor{midnightblue}{HTML}{191970} 
\newcommand{\ssl}[1]{\textcolor{midnightblue}{\small\scshape #1}}
\newcommand{\myparagraph}[1]{\noindent\textbf{#1} \hspace{.06cm}}

\title{\method{}: Gaussian-Regularized Speech SSL without Engineered Prediction Targets}
\name{\shortstack{Gaspard Bott\'{e}$^{1,2,\star}$, S\'{e}verin Baroudi$^{6}$, Samir Sadok$^{3}$, Francesco Paissan$^{2,5}$,\\ Thomas Hueber$^{7}$, Xavier Alameda-Pineda$^{3}$, Ricard Marxer$^{4,\star}$, Mirco Ravanelli$^{1,2,\star}$}\thanks{$^{\star}$Correspondence: \texttt{gaspard003.botte@gmail.com}, \texttt{ricard.marxer@lis-lab.fr}, \texttt{mirco.ravanelli@concordia.ca}}}
\address{$^{1}$Concordia University \quad $^{2}$Mila -- Qu\'{e}bec AI Institute \quad $^{4}$CNRS, ILLS, Univ Toulon\\
$^{3}$Inria, Univ. Grenoble Alpes, CNRS, LJK \quad $^{6}$Univ Toulon, Aix Marseille Univ, CNRS, LIS\\
$^{5}$Universit\'{e} Laval \quad $^{7}$Univ. Grenoble Alpes, CNRS, Grenoble INP, GIPSA-lab, Grenoble, France}

\begin{document}
\raggedbottom
\maketitle
\suppressfloats[t]

\begin{abstract}
Speech self-supervised learning aims to learn general-purpose representations for downstream speech tasks. However, current approaches rely on complex, carefully designed prediction targets.
We challenge this necessity with GLaS-JEPA, a framework that directly predicts the current encoder's continuous representations at masked positions, without contrastive learning, discrete targets, or separate EMA target encoders. We prevent representation collapse using SIGReg representation-space regularization, eliminating the need for engineered target-generation mechanisms. Pretrained on 960 hours of LibriSpeech, our 57M-parameter model achieves a 6.89\% WER on frozen-encoder SUPERB ASR and a 25.87\% CER on slot filling, outperforming the best non-distilled sub-90M baselines by 43.1\% and 22.0\%, respectively. These results demonstrate
that highly competitive speech representations can emerge from a radically simplified training recipe.
\end{abstract}
\begin{keywords}
self-supervised learning, speech representations, JEPA, Gaussian regularization
\end{keywords}

\section{Introduction}
\label{sec:intro}

Speech self-supervised learning (SSL) has emerged as a powerful paradigm, achieving state-of-the-art performance across diverse downstream tasks. By allowing a pretrained encoder to transfer effectively across recognition, speaker analysis, and spoken-language understanding, SSL demonstrates remarkable robustness, particularly in low-resource scenarios where labeled data is scarce \cite{baevski2020wav2vec,yang2021superb,chen2022wavlm}. This success motivates a closer look at the core ingredients required for compact, efficient encoders. A critical design dimension in masked-prediction speech SSL is target construction, which defines what the encoder learns to infer from context and directly shapes its learning signal. For instance, \ssl{Wav2vec~2.0} relies on quantized representations \cite{baevski2020wav2vec}, \ssl{HuBERT} and \ssl{WavLM} utilize clustered hidden units \cite{hsu2021hubert,chen2022wavlm}, \ssl{data2vec~2.0} leverages continuous targets from an exponential-moving-average (EMA) encoder \cite{baevski2023data2vec2}, and \ssl{S-JEPA} predicts soft Gaussian-mixture assignments \cite{ioannides2026sjepa}. These mechanisms do more than resist collapse: they fundamentally shape the information content and stability of the prediction signal. Simplified recipes such as S-JEPA still construct dedicated targets \cite{ioannides2026sjepa}. This raises a fundamental question: \emph{Are specially engineered prediction targets strictly necessary to achieve competitive performance in speech SSL?}

\begingroup
\setlength{\intextsep}{6pt}
\begin{figure}[t]
\centering
\def\OursSixLayerParams{43.22}
\def\OursSixLayerWER{8.75}

\begin{tikzpicture}
  \begin{axis}[
    scale only axis,
    width=7.82cm,
    height=3.65cm,
    xmin=0,xmax=102,
    ymin=3.5,ymax=22,
    xtick={0,20,40,60,80,100},
    ytick={5,10,15,20},
    xlabel={Parameters (M)},
    ylabel={WER (\%)},
    axis lines=left,
    axis line style={-{Latex[length=1.15mm,width=0.85mm]},black!82,line width=0.55pt},
    tick align=inside,
    major tick length=2pt,
    ticklabel shift=1pt,
    tick style={black!55,line width=0.45pt},
    grid=major,
    major grid style={draw=black!11,line width=0.35pt},
    label style={font=\footnotesize,inner sep=0pt},
    tick label style={font=\footnotesize,inner sep=0pt},
    xlabel style={at={(axis description cs:0.5,-0.09)},anchor=north},
    ylabel style={at={(axis description cs:-0.055,0.5)},anchor=south},
    every axis plot/.append style={line width=0.5pt},
    clip=false
  ]
    \addplot[only marks,mark=o,mark size=2.1pt,draw=black!55,fill=white]
      coordinates {(21.3,18.17) (32.54,15.86) (34.15,17.71) (89.84,13.02)};
    \addplot[only marks,mark=triangle*,mark size=2.2pt,
             draw=orange!75!black,fill=orange!75!black]
      coordinates {(51.8,12.10)};
    \addplot[only marks,mark=square,mark size=2.0pt,draw=black!58,fill=white]
      coordinates {(95.0,6.43)};
    \addplot[only marks,mark=triangle,mark size=2.1pt,draw=black!58,fill=white]
      coordinates {(94.7,6.42)};
    \addplot[only marks,mark=o,mark size=2.1pt,draw=black!75,fill=black!15]
      coordinates {(94.7,6.21)};

    \draw[densely dashed,black!52,line width=0.50pt]
      (axis cs:0,4.81) -- (axis cs:102,4.81);
    \addplot[only marks,mark=diamond*,mark size=2.2pt,
             draw=black!75,fill=black!40]
      coordinates {(93.78,4.81)};

    \addplot[mark=*,mark size=2.0pt,line width=0.45pt,draw=blue!30,
             mark options={draw=blue!55!black,fill=blue!62,line width=0.65pt}]
      coordinates {(29.07,12.24) (\OursSixLayerParams,\OursSixLayerWER) (57.36,6.89)};

    \node[font=\footnotesize,anchor=south east,fill=white,inner sep=1.0pt]
      at (axis cs:20.8,18.6) {TERA};
    \node[font=\footnotesize,anchor=south west,fill=white,inner sep=1.0pt]
      at (axis cs:35.3,18.1) {vq-wav2vec};
    \node[font=\footnotesize,anchor=north east,fill=white,inner sep=1.0pt]
      at (axis cs:31.0,15.5) {wav2vec};
    \node[font=\footnotesize,anchor=south east,fill=white,inner sep=1.0pt]
      at (axis cs:101,14.0) {DeCoAR 2.0};
    \node[font=\footnotesize,anchor=south,fill=white,inner sep=1.0pt]
      at (axis cs:51.8,12.85) {S-JEPA};
    \node[font=\footnotesize\bfseries,text=blue!55!black,anchor=south east,
          fill=white,inner sep=1.0pt]
      at (axis cs:41.0,8.75) {GLaS-JEPA};
    \node[font=\footnotesize,anchor=south east,align=right,fill=white,
          fill opacity=0.94,text opacity=1,inner sep=1.0pt]
      at (axis cs:97.5,7.35)
      {Base encoders\\(95M)};
  \end{axis}
\end{tikzpicture}

\caption{Frozen SUPERB ASR versus parameter count. All plotted models
are pretrained on LibriSpeech 960 h, except S-JEPA, pretrained on 83k h
\cite{yang2021superb,ioannides2026sjepa}.
Dashed line: data2vec~2.0 WER (4.81\%) \cite{yoon2023mcrdata2vec}.}
\label{fig:pareto}
\end{figure}
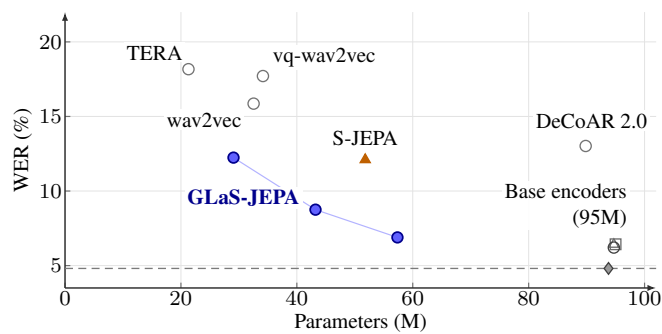
\endgroup

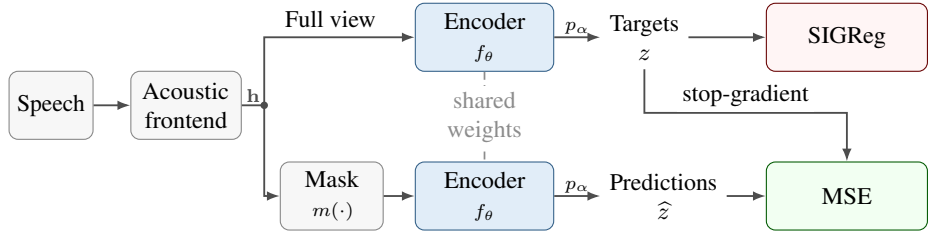
\begin{figure*}[t!]
\centering
\begin{tikzpicture}[
  x=0.94cm,y=1cm,font=\small,
  block/.style={draw=black!48,rounded corners=1.2mm,fill=black!3,
    minimum height=9mm,text width=1.42cm,align=center,inner sep=3pt},
  enc/.style={block,draw=MethodBorder,fill=MethodBlue,text width=1.62cm},
  loss/.style={block,text width=1.92cm},
  regloss/.style={loss,draw=red!40!black,fill=red!4},
  mseloss/.style={loss,draw=green!35!black,fill=green!5},
  flow/.style={-{Latex[length=1.7mm]},line width=0.7pt,draw=black!70}
]
  \node[block,text width=0.90cm] (audio) at (0.65,0.8) {Speech};
  \node[block,text width=1.25cm] (front) at (2.55,0.8) {Acoustic\\frontend};
  \coordinate (fork) at (3.65,0.8);
  \node[block,text width=1.15cm] (mask) at (4.6,-0.4) {Mask\\{\scriptsize $m(\cdot)$}};
  \node[enc] (full) at (6.75,1.7) {Encoder\\{\scriptsize $f_\theta$}};
  \node[enc] (masked) at (6.75,-0.4) {Encoder\\{\scriptsize $f_\theta$}};
  \node[align=center,font=\small] (target) at (9.0,1.7) {Targets\\$z$};
  \node[align=center,font=\small,anchor=west] (pred) at (target.west |- masked.center) {Predictions\\$\widehat z$};
  \node[regloss] (reg) at (11.85,1.7) {SIGReg};
  \node[mseloss] (mse) at (11.85,-0.4) {MSE};

  \draw[flow] (audio) -- (front);
  \draw[black!70,line width=0.7pt] (front.east) -- node[above,font=\scriptsize,inner sep=1pt] {$\mathbf h$} (fork);
  \fill[black!70] (fork) circle (1.4pt);
  \draw[flow] (fork) |- node[pos=0.72,above,font=\small] {Full view} (full.west);
  \draw[flow] (fork) |- (mask.west);
  \draw[flow] (mask) -- (masked);
  \draw[flow] (full) -- node[above,font=\scriptsize,inner sep=1pt] {$p_\alpha$} (target);
  \draw[flow] (masked) -- node[above,font=\scriptsize,inner sep=1pt] {$p_\alpha$} (pred);
  \draw[flow] (target) -- (reg);
  \draw[flow] (pred) -- (mse);
  \draw[flow] (target.south) -- (9.0,0.7) --
    node[above,font=\small,fill=white,inner sep=2pt] {stop-gradient}
    (11.85,0.7) -- (mse.north);
  \draw[densely dashed,black!45,line width=0.65pt]
    (full.south) -- node[fill=white,align=center,font=\small,inner sep=2pt]
    {shared\\weights} (masked.north);
\end{tikzpicture}
\caption{\method{}: shared encoder and token-wise projector $p_\alpha$. MSE uses masked predictions and stop-gradient targets; SIGReg backpropagates through the full view. The projector is discarded downstream.}
\label{fig:method}
\end{figure*}

To answer this, we introduce \ssl{\method{}} (\ssl{{G}}aussian \ssl{{La}}tent \ssl{{S}}peech-\ssl{{JEPA}}), a framework that directly predicts the current encoder's continuous representations at masked positions, bypassing discrete assignments or an EMA target encoder. Following the joint-embedding predictive architecture (JEPA) principle \cite{lecun2022ami,assran2023ijepa}, prediction occurs entirely in representation space, meaning targets evolve jointly with the model.
However, optimizing these unconstrained targets risks representation collapse, where the model minimizes prediction error by outputting constant vectors and discarding speech information. Rather than relying on a target generator to prevent this, \ssl{\method{}} uses Sketched Isotropic Gaussian Regularization (SIGReg) \cite{balestriero2025lejepa} to regularize the space directly. This shifts collapse prevention from complex target construction to a simple regularization term encouraging a non-collapsed Gaussian distribution. We evaluate our approach on the SUPERB benchmark \cite{yang2021superb}.

Our core contributions are: (1) introducing \ssl{\method{}}, a speech SSL framework that learns by predicting the encoder's continuous representations at masked positions; (2) showing that SIGReg is sufficient to prevent collapse in this setting, dispensing with the need for target-generation mechanisms commonly used in speech SSL; and (3) establishing strong performance across content, speaker, and semantic tasks, showing that competitive speech representations can emerge from a substantially simpler training pipeline.

\section{Related Work}
\myparagraph{Constructing prediction targets.}
Quantization supplies discrete latent targets in \ssl{wav2vec~2.0} and \ssl{vq-wav2vec} \cite{baevski2020wav2vec,baevski2020vqwav2vec}, whereas \ssl{HuBERT} and \ssl{WavLM} predict clustered pseudo-labels \cite{hsu2021hubert,chen2022wavlm}. BEST-RQ simplifies this process with a fixed random-projection quantizer \cite{chiu2022bestrq}, retaining a dedicated target mapping.
Continuous latent prediction is already established by \ssl{data2vec~2.0} \cite{baevski2023data2vec2}. It constructs targets by averaging normalized upper-layer representations from a slowly moving EMA teacher, and regresses them with a separate temporal CNN predictor. The distinction in \ssl{\method{}} is not continuity, but using current-encoder representations with explicit distributional regularization instead of processed EMA targets. Its token-wise linear projector maps latents independently into a 128-dimensional loss space; unlike the \ssl{data2vec~2.0} predictor, it does not mix neighboring positions.
\ssl{S-JEPA }\cite{ioannides2026sjepa} instead predicts soft GMM posteriors, using MFCC features initially and EMA-encoder features.  These alternatives motivate testing whether discrete target mappings and EMA-based target generation can be replaced by direct regularization of current-encoder representations.

\myparagraph{Regularizing representations directly.}
VICReg and Barlow Twins explicitly constrain representation statistics, providing an alternative to predictor and stop-gradient mechanisms such as those used in SimSiam\cite{bardes2022vicreg,zbontar2021barlow,chen2021simsiam}. 
\ssl{LeJEPA} introduces SIGReg for distributional regularization \cite{balestriero2025lejepa}, and LeWorldModel applies it to visual trajectories \cite{maes2026leworldmodel}. Building on SIGReg, we investigate whether direct representation regularization can enable current-encoder speech latent prediction without discrete assignments or an EMA teacher. Speech produces many latent frames per utterance, yielding large token populations within each batch; we study how to group them in Sec.~\ref{sec:distributional-regularization}.

\myparagraph{Scope of comparison.}
Related non-distilled speech encoders include modified \ssl{CPC}, \ssl{APC}, \ssl{PASE+}, \ssl{TERA}, \ssl{wav2vec}, and \ssl{DeCoAR~2.0} \cite{riviere2020cpc,chung2019apc,ravanelli2020paseplus,liu2021tera,schneider2019wav2vec,ling2020decoar2}, covering contrastive prediction, acoustic reconstruction, and multi-task objectives. \ssl{A-JEPA}, \ssl{Audio-JEPA}, and \ssl{WavJEPA} explore related predictive representations for general audio \cite{fei2023ajepa,tuncay2025audiojepa,yuksel2025wavjepa}; we discuss them but exclude unmatched numerical comparisons because their general-audio training and evaluation differ from our speech-only setting.

\section{\method}
\label{sec:method}

\begin{table*}[t!]
\centering
\caption{Frozen-encoder SUPERB results (\%). Baselines: SUPERB/S-JEPA \cite{yang2021superb,ioannides2026sjepa}; SD: WavLM \cite{chen2022wavlm}; data2vec~2.0: ASR/ER/SF \cite{yoon2023mcrdata2vec}, SD \cite{chang2025usad}. Bold: best non-distilled sub-90M score. \method{} uses Full Marginal SIGReg. Parameters rounded to millions.}
\label{tab:superb}
\small
\setlength{\tabcolsep}{5pt}
\begin{tabular}{lrrrrrrr}
\toprule
& & & \multicolumn{1}{c}{Content} & \multicolumn{1}{c}{Speaker} & \multicolumn{1}{c}{Paralinguistic} & \multicolumn{2}{c}{Semantic}\\
\cmidrule(lr){4-4}\cmidrule(lr){5-5}\cmidrule(lr){6-6}\cmidrule(lr){7-8}
Model & Params (M) & Pretrain (h) & ASR WER $\downarrow$ & SD DER $\downarrow$ & ER Acc. $\uparrow$ & SF F1 $\uparrow$ & SF CER $\downarrow$\\
\midrule
\multicolumn{8}{l}{\textit{Base encoders (approximately 95M parameters)}}\\
wav2vec~2.0 Base & 95 & 960 & 6.43\% & 6.08 & 63.43 & 88.30 & 24.77\\
HuBERT Base & 95 & 960 & 6.42\% & 5.88 & 64.92 & 88.53 & 25.20\\
WavLM Base & 95 & 960 & 6.21\% & 4.55 & 65.94 & 89.38 & 22.86\\
data2vec~2.0 Base & 94 & 960 & 4.81\% & 6.5 & 66.66 & 89.67 & 22.09\\
\midrule
\multicolumn{8}{l}{\textit{Non-distilled sub-90M baselines}}\\
TERA & 21 & 960 & 18.17\% & 9.96 & 56.27 & 67.50 & 54.17\\
wav2vec & 33 & 960 & 15.86\% & 9.90 & 59.79 & 76.37 & 43.71\\
vq-wav2vec & 34 & 960 & 17.71\% & 9.93 & 58.24 & 77.68 & 41.54\\
DeCoAR~2.0 & 90 & 960 & 13.02\% & 6.59 & 62.47 & 83.28 & 34.73\\
S-JEPA & 52 & 83k & 12.10\% & -- & \textbf{64.83} & 83.05 & 33.17\\
\midrule
\rowcolor{OursGreen}\textbf{\method{}} & 57 & 960 & \textbf{6.89\%} & \textbf{6.46} & 57.91 & \textbf{87.72} & \textbf{25.87}\\
\bottomrule
\end{tabular}
\vspace{-10pt}
\end{table*}

\subsection{Architecture and forward pass}
We introduce \textbf{GLaS-JEPA} (\textbf{G}aussian \textbf{La}tent \textbf{S}peech Joint-Embedding Predictive Architecture), a dual-path framework for direct prediction of continuous speech representations. As illustrated in Fig.~\ref{fig:method}, an acoustic frontend first processes the input speech into a sequence of continuous feature vectors $\mathbf{h}$. This sequence is then branched into two parallel processing pathways: The full view provides the complete, uncorrupted feature sequence $\mathbf{h}$ to the encoder $f_\theta$, and a subsequent projector $p_\alpha$ maps each token independently into a dedicated loss space to target representations: $\mathbf{z} = p_\alpha\big(f_\theta(\mathbf{h})\big).$
The masked view applies a masking function $m(\cdot)$ that replaces contiguous spans of the input features with a learnable mask token. The same encoder $f_\theta$ processes these features and learns to infer missing content from temporal context. The shared projector $p_\alpha$ then maps its representations into the loss space: $\widehat{\mathbf{z}} = p_\alpha\big(f_\theta\big(m(\mathbf{h})\big)\big).
$

The encoder learns prediction during training; there is no separate temporal predictor or EMA teacher. The token-wise linear projector provides a 128-dimensional bottleneck that separates the optimization loss space from the encoder representations used downstream. It can assist optimization but does not mix neighboring positions, and is discarded after pretraining.

\subsection{Loss function and regularization}

\myparagraph{Prediction.}
The model learns by minimizing the mean squared error (MSE) between the masked-view predictions $\widehat{\mathbf{z}}$ and the full-view targets $\mathbf{z}$. We simplify the notation by letting
$\mathcal{M}\subseteq\{1,\ldots,B\}\times\{1,\ldots,T\},$
represent the set of all masked time indices across a batch. The prediction loss is computed exclusively at these masked positions:
\begin{equation}
    \mathcal{L}_{\mathrm{pred}}
    =
    \frac{1}{|\mathcal{M}|}
    \sum_{(b,t)\in\mathcal{M}}
    \left\|
    \widehat{\mathbf{z}}_{b,t}
    -
    \sg(\mathbf{z}_{b,t})
    \right\|_2^2.
    \label{eq:mse}
\end{equation}
where $\sg(\cdot)$ denotes the stop-gradient operation. Because the targets $\mathbf{z}$ are generated dynamically using the current, actively updating network weights, the stop-gradient ensures the prediction loss routes gradients only through the masked forward path.

\myparagraph{Regularization.}
SIGReg \cite{balestriero2025lejepa} regularizes unmasked targets $z$ to prevent collapse. It averages weighted squared discrepancies between empirical characteristic functions of random unit projections and the standard Gaussian function $e^{-u^2/2}$ at frequency $u$, encouraging isotropic Gaussian representations. We optimize:
\begin{equation}
    \mathcal{L} = (1-\lambda)\mathcal{L}_{\mathrm{pred}} + \lambda \mathcal{L}_{\mathrm{reg}}, \quad \text{with } \lambda = 0.01.
\end{equation}
Klindt et al.\ \cite{klindt2026worldmodel} prove that LeJEPA guarantees linear identifiability under independent Gaussian latents and isotropic transitions, a property unique to Gaussian distributions. This raises a key empirical question for speech: \emph{is Gaussian regularization sufficient to extract useful, non-collapsed representations from complex, correlated audio?} \method{} investigates this by shifting collapse prevention entirely to $\mathcal{L}_{\mathrm{reg}}$. Here, gradients from $\mathcal{L}_{\mathrm{reg}}$ flow exclusively through the unmasked full-view representations, preventing representation collapse without strictly enforcing exact Gaussianity on finite-length speech latents.

\subsection{Applying Distributional Regularization}
\label{sec:distributional-regularization}
Speech representations are temporally correlated. For a latent batch $Z\in\mathbb R^{B\times T\times d_p}$, we must choose which population SIGReg should regularize. We consider two alternatives (Fig.~\ref{fig:populations}).

\begin{figure}[!b]
\centering
\resizebox{\columnwidth}{!}{
\begingroup
\definecolor{PopBlue}{RGB}{66,116,173}
\definecolor{PopOrange}{RGB}{207,145,65}
\definecolor{PopGreen}{RGB}{69,143,112}
\newcommand{\PopToken}[4]{%
  \def\popcolor{PopBlue}%
  \if B#1\def\popcolor{PopOrange}\fi
  \if C#1\def\popcolor{PopGreen}\fi
  \path[draw=\popcolor!80!black,fill=\popcolor!20,rounded corners=.5pt]
    (#3,#4) rectangle ++(.48,.34);
  \node[font=\scriptsize,inner sep=0pt] at (#3+.24,#4+.17) {$#1_{#2}$};
}
\begin{tikzpicture}[x=1cm,y=1cm,line width=.35pt]
  \node[font=\scriptsize\bfseries] at (1.65,2.22) {(a) Time-conditional};
  \node[font=\scriptsize\bfseries] at (5.7,2.22) {(b) Shuffled Marginal};
  \node[font=\scriptsize] at (1.65,1.87) {$p(z\mid t)$: fixed index};
  \node[font=\scriptsize] at (5.7,1.87) {$p(z)$: mixed indices};
  \foreach \t in {1,...,4}{
    \pgfmathsetmacro{\xx}{.5+(\t-1)*.58}
    \PopToken{A}{\t}{\xx}{1.29}
    \PopToken{B}{\t}{\xx}{.85}
    \PopToken{C}{\t}{\xx}{.41}
  }
  \PopToken{A}{1}{4.55}{1.29}
  \PopToken{C}{3}{4.55}{.85}
  \PopToken{A}{4}{4.55}{.41}
  \PopToken{B}{2}{5.13}{1.29}
  \PopToken{A}{3}{5.13}{.85}
  \PopToken{C}{1}{5.13}{.41}
  \PopToken{C}{4}{5.71}{1.29}
  \PopToken{B}{1}{5.71}{.85}
  \PopToken{B}{4}{5.71}{.41}
  \PopToken{A}{2}{6.29}{1.29}
  \PopToken{C}{2}{6.29}{.85}
  \PopToken{B}{3}{6.29}{.41}
  \foreach \xx in {1.08,4.55}{
    \draw[red!65!black,line width=.9pt,rounded corners=1pt]
      (\xx-.05,.36) rectangle (\xx+.53,1.68);
  }
  \node[font=\scriptsize] at (3.65,-.03)
    {Population sample $\rightarrow$ SIGReg $\rightarrow$ $\mathcal N(0,I)$ reference};
\end{tikzpicture}
\endgroup}
\par\vspace{4pt}
\begingroup
\small
\setlength{\tabcolsep}{4pt}
\begin{tabular}{lrrr}
\toprule
SIGReg & ASR WER $\downarrow$ & ER Acc. $\uparrow$ & SD DER $\downarrow$\\
\midrule
Time-conditional & 13.99 & 57.92 & 7.65\\
Shuffled Marginal & 12.24 & 58.60 & 6.65\\
\bottomrule
\end{tabular}
\par\endgroup
\caption{Time-conditional (left) and Shuffled Marginal (right), with SUPERB results for the 30M model. Colors/letters: utterances; subscripts: time indices. Outlines mark groups of $B$ tokens.}
\label{fig:populations}
\end{figure}

\myparagraph{Time-conditional SIGReg.}
This regularizes $p(z\mid t)$. At each fixed latent index $t$, the population contains $B$ representations from different utterances. We average the $T$ penalties, as in LeWorldModel \cite{maes2026leworldmodel}. Using distinct utterances limits within-utterance dependence, although shared-speaker dependence may remain.

\myparagraph{Marginal SIGReg.}
This regularizes the batch-induced marginal $p(z)$, ignoring the time index. Tokens across times and utterances belong to the same population, allowing both sources of variation to maintain a non-collapsed signal. We use two estimators of this marginal objective.

\emph{Shuffled Marginal}, used only in the controlled four-layer ablation, shuffles the $BT$ tokens into $T$ groups of $B$ and averages their penalties. This matches Time-conditional SIGReg in group size, number of penalties, and computation, isolating population construction rather than sample size.
\emph{Full Marginal}, used in our main 57M model, applies one penalty to all $BT$ tokens. Both target the same marginal population $p(z)$ and differ only in finite-sample estimation. Under IID sampling, both converge to the same population-level objective as sample size grows; speech tokens need not satisfy this assumption.

With all other settings fixed, Shuffled Marginal improves ASR WER from 13.99\% to 12.24\%, ER accuracy from 57.92\% to 58.60\%, and SD DER from 7.65\% to 6.65\% (Fig.~\ref{fig:populations}). These gains motivate Marginal SIGReg for the main model.

\section{Experimental Setup}
\textbf{Encoder and training.}
We pretrain on LibriSpeech 960 h \cite{panayotov2015librispeech}.
The eight-layer Conformer \cite{gulati2020conformer} has width 576, eight
heads, FFN width 2,048, and convolution kernel 31. The frontend uses
80-bin log-Mel features (25 ms window, 5 ms hop) with temporal stride four,
yielding 20 ms latent spacing. A token-wise linear projector maps encoder
representations to a 128-dimensional loss space. The downstream encoder
contains approximately 57M parameters, excluding the projector.

We mask 50\% of positions in contiguous spans of ten latents (200 ms).
Training uses 220,000 AdamW updates, 40,000-step linear warmup followed by
cosine decay, peak learning rate $10^{-4}$, weight decay $10^{-3}$, and
$\lambda=0.01$. Crops span 2--15.6 s, with approximately 4,000 s of audio
per optimization step. We use Full Marginal SIGReg on all tokens in the
batch.

\textbf{Evaluation.}
For the four SUPERB tasks \cite{yang2021superb}, we freeze the encoder and train task heads over a learned weighted sum of hidden layers. We follow WavLM Base learning
rates and batch sizes \cite{chen2022wavlm} for ASR, speaker diarization (SD),
emotion recognition (ER), and slot filling (SF); ASR uses no external
language model. Published baselines are contextual comparisons, not
matched-budget reimplementations. Linear probes assess 41-phone
classification using CPC's LibriSpeech protocol and speaker identification
from time-pooled features; RankMe \cite{garrido2023rankme} measures
effective rank.

\section{Results and Discussion}

\subsection{Downstream transfer}
Table~\ref{tab:superb} tests whether current-representation prediction
supports useful transfer. \method{} achieves 6.89\% WER and 25.87\% SF
concept error rate (CER), improving over the best listed non-distilled
sub-90M baselines by 43.1\% and 22.0\% relatively. Against S-JEPA, it uses
57.36M rather than 51.8M parameters and 960 rather than approximately
83,000 pretraining hours. Baseline comparisons use different architectures and training budgets, so they cannot isolate the SSL objective's effect.
Accordingly, these comparisons test whether such a target-free model can be competitive, rather than whether its objective outperforms alternatives under controlled conditions.

The gains are task-dependent: SF F1 reaches 87.72\%, but ER accuracy
(57.91\%) trails S-JEPA (64.83\%). \method{} reaches
6.46\% SD DER, slightly below DeCoAR~2.0 (6.59\%) and data2vec~2.0 (6.5\%), with approximately 39\% fewer parameters. S-JEPA does not report SD here. Base encoders remain
stronger in ASR, including WavLM (6.21\% WER) and data2vec~2.0 (4.81\%).

\subsection{Collapse, representations, and limitations}
\begin{figure}[!t]
\centering
\includegraphics[width=\columnwidth,trim=13.32bp 3.06bp 4.32bp 6.84bp,clip]{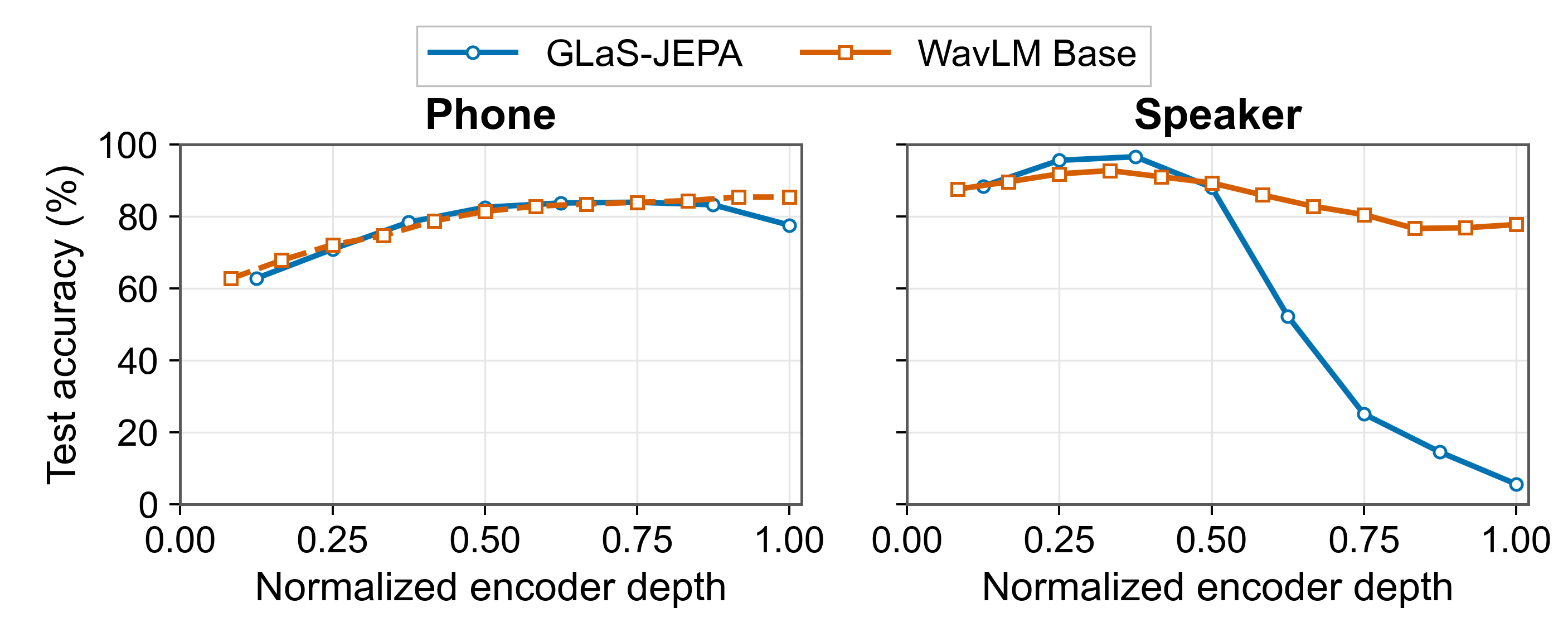}
\caption{Phone and speaker linear-probe test accuracy versus normalized encoder depth for \method{} and WavLM Base.}
\label{fig:layerwise}
\end{figure}

Without SIGReg, effective rank \cite{garrido2023rankme} falls to 1 within a few thousand updates; phone accuracy is $\sim$16\% with imbalanced classes. Fig.~\ref{fig:layerwise} tracks phone and speaker linear-probe test accuracy across encoder layers, indicating where each type of information is accessible. Depth is normalized by the number of layers to compare the 8-layer \method{} and 12-layer WavLM Base. Their phone peaks are 83.9/85.4\% at layers 6/8 and 11/12; speaker peaks are 96.5/92.7\% at 3/8 and 4/12. Strong intermediate speaker encoding, then greater upper-layer loss than WavLM with retained phonetics, suggests invariance to slowly varying/global acoustic factors relevant to ER. From $\sim$140k to 220k steps in the same run, ASR WER improved from 7.52\% to 6.89\%, but ER fell from 59.20\% to 57.91\%. This suggests optimization increasingly favors content over paralinguistic information, though it does not establish causality.

Stable current-encoder training at $\sim$95M remains ongoing work; 57M competitiveness does not establish equivalent scaling to Base WavLM, HuBERT, or data2vec~2.0.

\section{Conclusion}
At the demonstrated scale, GLaS-JEPA replaces engineered targets with representation regularization, reaching 6.89\% ASR WER at 57M parameters with competitive content, speaker, and semantic transfer. Competitive speech representations can thus emerge from current-encoder prediction without dedicated target generation.

\clearpage
\begingroup
\fontsize{9}{11.3}\selectfont
\noindent\textbf{Acknowledgments.}
We thank Dr.~Thomas Hueber (GIPSA-lab, CNRS) for making this internship project possible. We thank Lovanya Jain for help with experiments. M.~Ravanelli acknowledges support from NSERC, the Digital Research Alliance of Canada (\url{alliancecan.ca}), and Translated for funding through the Immediate Research Grant. This work benefited from support from the French National Research Agency through the ANR-20-CE23-0012-01 (MIM) grant and from access to the HPC resources of IDRIS under allocation A0191014044.
\par

\let\originalthebibliography\thebibliography
\renewcommand{\thebibliography}[1]{\originalthebibliography{#1}\setlength{\itemsep}{0pt}}
\interlinepenalty=10000
\tolerance=1000
\emergencystretch=1em
\bibliographystyle{IEEEbib}
\bibliography{references}

@misc{lecun2022ami,
  author = {LeCun, Yann},
  title = {A Path Towards Autonomous Machine Intelligence},
  howpublished = {Position paper, OpenReview},
  note = {Version 0.9.2},
  year = {2022},
  url = {https://openreview.net/forum?id=BZ5a1r-kVsf}
}

@inproceedings{schneider2019wav2vec,
  author = {Schneider, Steffen and Baevski, Alexei and Collobert, Ronan and Auli, Michael},
  title = {{wav2vec}: Unsupervised Pre-Training for Speech Recognition},
  booktitle = {Interspeech},
  pages = {3465--3469},
  doi = {10.21437/Interspeech.2019-1873},
  year = {2019},
  url = {https://www.isca-archive.org/interspeech_2019/schneider19_interspeech.html}
}

@inproceedings{baevski2020vqwav2vec,
  author = {Baevski, Alexei and Schneider, Steffen and Auli, Michael},
  title = {{vq-wav2vec}: Self-Supervised Learning of Discrete Speech Representations},
  booktitle = {ICLR},
  year = {2020},
  url = {https://arxiv.org/abs/1910.05453}
}

@article{ling2020decoar2,
  author = {Ling, Shaoshi and Liu, Yuzong},
  title = {{DeCoAR} 2.0: Deep Contextualized Acoustic Representations with Vector Quantization},
  journal = {arXiv preprint arXiv:2012.06659},
  year = {2020},
  url = {https://arxiv.org/abs/2012.06659}
}

@article{yuksel2025wavjepa,
  author = {Yuksel, Goksenin and Guetschel, Pierre and Tangermann, Michael and van Gerven, Marcel and van der Heijden, Kiki},
  title = {{WavJEPA}: Semantic Learning Unlocks Robust Audio Foundation Models for Raw Waveforms},
  journal = {arXiv preprint arXiv:2509.23238},
  year = {2025},
  url = {https://arxiv.org/abs/2509.23238}
}

@article{baevski2020wav2vec,
  author  = {Baevski, Alexei and Zhou, Henry and Mohamed, Abdelrahman and Auli, Michael},
  title   = {wav2vec 2.0: A Framework for Self-Supervised Learning of Speech Representations},
  journal = {NeurIPS},
  volume  = {33},
  pages   = {12449--12460},
  year    = {2020},
  url = {https://arxiv.org/abs/2006.11477}
}

@article{hsu2021hubert,
  author  = {Hsu, Wei-Ning and Bolte, Benjamin and Tsai, Yao-Hung Hubert and Lakhotia, Kushal and Salakhutdinov, Ruslan and Mohamed, Abdelrahman},
  title   = {{HuBERT}: Self-Supervised Speech Representation Learning by Masked Prediction of Hidden Units},
  journal = {IEEE/ACM Trans. Audio Speech Lang. Process.},
  volume  = {29},
  pages   = {3451--3460},
  year    = {2021},
  url = {https://arxiv.org/abs/2106.07447}
}

@article{chen2022wavlm,
  author  = {Chen, Sanyuan and Wang, Chengyi and Chen, Zhengyang and Wu, Yu and Liu, Shujie and Chen, Zhuo and Li, Jinyu and Kanda, Naoyuki and Yoshioka, Takuya and Xiao, Xiong and Wu, Jian and Zhou, Long and Ren, Shuo and Qian, Yanmin and Qian, Yao and Wu, Jian and Zeng, Michael and Yu, Xiangzhan and Wei, Furu},
  title   = {{WavLM}: Large-Scale Self-Supervised Pre-Training for Full Stack Speech Processing},
  journal = {IEEE J. Sel. Topics Signal Process.},
  volume  = {16},
  number  = {6},
  pages   = {1505--1518},
  year    = {2022},
  url = {https://doi.org/10.1109/JSTSP.2022.3188113}
}

@inproceedings{baevski2023data2vec2,
  author    = {Baevski, Alexei and Babu, Arun and Hsu, Wei-Ning and Auli, Michael},
  title     = {Efficient Self-Supervised Learning with Contextualized Target Representations for Vision, Speech and Language},
  booktitle = {ICML},
  series = {PMLR},
  volume    = {202},
  pages     = {1416--1429},
  year      = {2023},
  url = {https://proceedings.mlr.press/v202/baevski23a.html}
}

@inproceedings{assran2023ijepa,
  author    = {Assran, Mahmoud and Duval, Quentin and Misra, Ishan and Bojanowski, Piotr and Vincent, Pascal and Rabbat, Michael and LeCun, Yann and Ballas, Nicolas},
  title     = {Self-Supervised Learning from Images with a Joint-Embedding Predictive Architecture},
  booktitle = {CVPR},
  pages     = {15619--15629},
  year      = {2023},
  url = {https://arxiv.org/abs/2301.08243}
}

@article{balestriero2025lejepa,
  journal       = {arXiv preprint arXiv:2511.08544},
  author        = {Balestriero, Randall and LeCun, Yann},
  title         = {{LeJEPA}: Provable and Scalable Self-Supervised Learning Without the Heuristics},
  year          = {2025},
  eprint        = {2511.08544},
  archivePrefix = {arXiv},
  primaryClass  = {cs.LG},
  url           = {https://arxiv.org/abs/2511.08544}
}

@article{maes2026leworldmodel,
  journal       = {arXiv preprint arXiv:2603.19312},
  author        = {Maes, Lucas and Le Lidec, Quentin and Scieur, Damien and LeCun, Yann and Balestriero, Randall},
  title         = {{LeWorldModel}: Stable End-to-End Joint-Embedding Predictive Architecture from Pixels},
  year          = {2026},
  eprint        = {2603.19312},
  archivePrefix = {arXiv},
  primaryClass  = {cs.LG},
  url           = {https://arxiv.org/abs/2603.19312}
}

@article{fei2023ajepa,
  journal       = {arXiv preprint arXiv:2311.15830},
  author        = {Fei, Zhengcong and Fan, Mingyuan and Huang, Junshi},
  title         = {{A-JEPA}: Joint-Embedding Predictive Architecture Can Listen},
  year          = {2023},
  eprint        = {2311.15830},
  archivePrefix = {arXiv},
  primaryClass  = {cs.SD},
  url           = {https://arxiv.org/abs/2311.15830}
}

@inproceedings{tuncay2025audiojepa,
  booktitle     = {ICME},
  author        = {Tuncay, Ludovic and Labb{\'e}, Etienne and Benetos, Emmanouil and Pellegrini, Thomas},
  title         = {{Audio-JEPA}: Joint-Embedding Predictive Architecture for Audio Representation Learning},
  year          = {2025},
  eprint        = {2507.02915},
  archivePrefix = {arXiv},
  primaryClass  = {cs.SD},
  url           = {https://arxiv.org/abs/2507.02915}
}

@article{ioannides2026sjepa,
  journal       = {arXiv preprint arXiv:2606.19398},
  author        = {Ioannides, Georgios and Kieback, Adrian and Goldfeder, Judah and Pang, Linsey and Chadha, Aman and Elkins, Aaron and LeCun, Yann and Shwartz-Ziv, Ravid},
  title         = {{S-JEPA}: Soft Clustering Anchors for Self-Supervised Speech Representation Learning},
  year          = {2026},
  eprint        = {2606.19398},
  archivePrefix = {arXiv},
  primaryClass  = {cs.SD},
  url           = {https://arxiv.org/abs/2606.19398}
}

@inproceedings{gulati2020conformer,
  author    = {Gulati, Anmol and Qin, James and Chiu, Chung-Cheng and Parmar, Niki and Zhang, Yu and Yu, Jiahui and Han, Wei and Wang, Shibo and Zhang, Zhengdong and Wu, Yonghui and Pang, Ruoming},
  title     = {Conformer: Convolution-Augmented Transformer for Speech Recognition},
  booktitle = {Interspeech},
  pages     = {5036--5040},
  year      = {2020},
  url = {https://www.isca-archive.org/interspeech_2020/gulati20_interspeech.html}
}

@inproceedings{panayotov2015librispeech,
  author    = {Panayotov, Vassil and Chen, Guoguo and Povey, Daniel and Khudanpur, Sanjeev},
  title     = {{LibriSpeech}: An {ASR} Corpus Based on Public Domain Audio Books},
  booktitle = {ICASSP},
  pages     = {5206--5210},
  year      = {2015},
  url = {https://doi.org/10.1109/ICASSP.2015.7178964}
}

@inproceedings{yang2021superb,
  booktitle     = {Interspeech},
  pages         = {1194--1198},
  doi           = {10.21437/Interspeech.2021-1775},
  author        = {Yang, Shu-wen and Chi, Po-Han and Chuang, Yung-Sung and Lai, Cheng-I Jeff and Lakhotia, Kushal and Lin, Yist Y. and Liu, Andy T. and Shi, Jiatong and Chang, Xuankai and Lin, Guan-Ting and Huang, Tzu-Hsien and Tseng, Wei-Cheng and Lee, Ko-tik and Liu, Da-Rong and Huang, Zili and Dong, Shuyan and Li, Shang-Wen and Watanabe, Shinji and Mohamed, Abdelrahman and Lee, Hung-yi},
  title         = {{SUPERB}: Speech Processing Universal PERformance Benchmark},
  year          = {2021},
  eprint        = {2105.01051},
  archivePrefix = {arXiv},
  primaryClass  = {cs.CL},
  url           = {https://arxiv.org/abs/2105.01051}
}

@inproceedings{garrido2023rankme,
  author    = {Garrido, Quentin and Balestriero, Randall and Najman, Laurent and LeCun, Yann},
  title     = {{RankMe}: Assessing the Downstream Performance of Pretrained Self-Supervised Representations by Their Rank},
  booktitle = {ICML},
  series = {PMLR},
  volume    = {202},
  pages     = {10929--10974},
  year      = {2023},
  url = {https://proceedings.mlr.press/v202/garrido23a.html}
}

@inproceedings{riviere2020cpc,
  author = {Rivi{\`e}re, Morgane and Joulin, Armand and Mazar{\'e}, Pierre-Emmanuel and Dupoux, Emmanuel},
  title = {Unsupervised Pretraining Transfers Well Across Languages},
  booktitle = {ICASSP},
  year = {2020},
  url = {https://arxiv.org/abs/2002.02848}
}

@inproceedings{chung2019apc,
  author = {Chung, Yu-An and Hsu, Wei-Ning and Tang, Hao and Glass, James},
  title = {An Unsupervised Autoregressive Model for Speech Representation Learning},
  booktitle = {Interspeech},
  pages = {146--150},
  year = {2019},
  doi = {10.21437/Interspeech.2019-1473},
  url = {https://www.isca-archive.org/interspeech_2019/chung19_interspeech.html}
}

@inproceedings{ravanelli2020paseplus,
  author = {Ravanelli, Mirco and Zhong, Jianyuan and Pascual, Santiago and Swietojanski, Pawel and Monteiro, Joao and Trmal, Jan and Bengio, Yoshua},
  title = {Multi-Task Self-Supervised Learning for Robust Speech Recognition},
  booktitle = {ICASSP},
  year = {2020},
  url = {https://arxiv.org/abs/2001.09239}
}

@article{liu2021tera,
  author = {Liu, Andy T. and Li, Shang-Wen and Lee, Hung-yi},
  title = {{TERA}: Self-Supervised Learning of Transformer Encoder Representation for Speech},
  journal = {IEEE/ACM Trans. Audio Speech Lang. Process.},
  volume = {29},
  year = {2021},
  doi = {10.1109/TASLP.2021.3095662},
  url = {https://arxiv.org/abs/2007.06028}
}

@inproceedings{chen2021simsiam,
  author = {Chen, Xinlei and He, Kaiming},
  title = {Exploring Simple Siamese Representation Learning},
  booktitle = {CVPR},
  pages = {15750--15758},
  year = {2021},
  url = {https://openaccess.thecvf.com/content/CVPR2021/html/Chen_Exploring_Simple_Siamese_Representation_Learning_CVPR_2021_paper.html}
}

@inproceedings{bardes2022vicreg,
  author = {Bardes, Adrien and Ponce, Jean and LeCun, Yann},
  title = {{VICReg}: Variance-Invariance-Covariance Regularization for Self-Supervised Learning},
  booktitle = {ICLR},
  year = {2022},
  url = {https://openreview.net/forum?id=xm6YD62D1Ub}
}

@inproceedings{zbontar2021barlow,
  author = {Zbontar, Jure and Jing, Li and Misra, Ishan and LeCun, Yann and Deny, St{\'e}phane},
  title = {Barlow Twins: Self-Supervised Learning via Redundancy Reduction},
  booktitle = {ICML},
  series = {PMLR},
  volume = {139},
  pages = {12310--12320},
  year = {2021},
  url = {https://proceedings.mlr.press/v139/zbontar21a.html}
}

@inproceedings{chiu2022bestrq,
  author = {Chiu, Chung-Cheng and Qin, James and Zhang, Yu and Yu, Jiahui and Wu, Yonghui},
  title = {Self-Supervised Learning with Random-Projection Quantizer for Speech Recognition},
  booktitle = {ICML},
  series = {PMLR},
  volume = {162},
  pages = {3915--3924},
  year = {2022},
  url = {https://proceedings.mlr.press/v162/chiu22a.html}
}

@inproceedings{yoon2023mcrdata2vec,
  author = {Yoon, Ji Won and Kim, Seok Min and Kim, Nam Soo},
  title = {{MCR-Data2vec 2.0}: Improving Self-supervised Speech Pre-training via Model-level Consistency Regularization},
  booktitle = {Interspeech},
  year = {2023},
  pages = {2833--2837},
  doi = {10.21437/Interspeech.2023-1579},
  url = {https://www.isca-archive.org/interspeech_2023/yoon23c_interspeech.html}
}

@article{klindt2026worldmodel,
  author = {Klindt, David and LeCun, Yann and Balestriero, Randall},
  title = {When Does {LeJEPA} Learn a World Model?},
  journal = {arXiv preprint arXiv:2605.26379},
  year = {2026},
  url = {https://arxiv.org/abs/2605.26379}
}

@inproceedings{chang2025usad,
  author = {Chang, Heng-Jui and Bhati, Saurabhchand and Glass, James and Liu, Alexander H.},
  title = {{USAD}: Universal Speech and Audio Representation via Distillation},
  booktitle = {IEEE ASRU},
  year = {2025},
  url = {https://sls.csail.mit.edu/publications/2025/HChang_ASRU-2025.pdf}
}
\endgroup
\end{document}